\documentclass[10pt,aps,prb,superscriptaddress,twocolumn,floatfix]{revtex4-1}
\usepackage{graphicx}
\usepackage{dcolumn}
\usepackage{bm}
\usepackage{amsmath}
\usepackage{subfigure}
\usepackage{textcomp}

\begin{document}

\newcommand{\eps}{$\varepsilon$-Fe$_2$O$_3$}
\newcommand{\gam}{$\gamma$-Fe$_2$O$_3$}

\title{Stabilization of \eps{} epitaxial layer on MgO(111)/GaN via an intermediate $\gamma$-phase}

\author{Victor Ukleev}
\affiliation{Laboratory for Neutron Scattering and Imaging (LNS), Paul Scherrer Institute (PSI), CH-5232 Villigen, Switzerland}
\email{victor.ukleev@psi.ch}
\author{Mikhail Volkov}
\affiliation{Ioffe Institute, 194021 Saint-Petersburg, Russia}
\author{Alexander Korovin}
\affiliation{Ioffe Institute, 194021 Saint-Petersburg, Russia}
\author{Thomas Saerbeck}
\affiliation{Institut Laue-Langevin, 71 Avenue des Martyrs, 38042 Grenoble, France}
\author{Nikolai Sokolov}
\affiliation{Ioffe Institute, 194021 Saint-Petersburg, Russia}
\author{Sergey Suturin}
\affiliation{Ioffe Institute, 194021 Saint-Petersburg, Russia}
\email{suturin@mail.ioffe.ru}

\begin{abstract}
In the present study we have demonstrated epitaxial stabilization of the metastable magnetically-hard $\varepsilon$-Fe$_2$O$_3$ phase on top of a thin MgO(111) buffer layer grown onto the GaN (0001) surface. The primary purpose to introduce a 4\,nm-thick buffer layer of MgO in between Fe$_2$O$_3$ and GaN was to stop thermal migration of Ga into the iron oxide layer. Though such migration and successive formation of the orthorhombic GaFeO$_3$ was supposed earlier to be a potential trigger of the nucleation of the isostructural $\varepsilon$-Fe$_2$O$_3$, the present work demonstrates that the growth of single crystalline uniform films of epsilon ferrite by pulsed laser deposition is possible even on the MgO capped GaN. The structural properties of the 60\,nm thick Fe$_2$O$_3$ layer on MgO / GaN  were probed by electron and x-ray diffraction, both suggesting that the growth of $\varepsilon$-Fe$_2$O$_3$ is preceded by formation of a thin layer of $\gamma$-Fe$_2$O$_3$. The presence of the magnetically hard epsilon ferrite was independently confirmed by temperature dependent magnetometry measurements. The depth-resolved x-ray and polarized neutron reflectometry reveal that the 10\,nm iron oxide layer at the interface has a lower density and a higher magnetization than the main volume of the $\varepsilon$-Fe$_2$O$_3$ film. The density and magnetic moment depth profiles derived from fitting the reflectometry data are in a good agreement with the presence of the magnetically degraded $\gamma$-Fe$_2$O$_3$ transition layer between MgO and $\varepsilon$-Fe$_2$O$_3$. The natural occurrence of the interface between magnetoelectric $\varepsilon$- and spin caloritronic $\gamma$- iron oxide phases can enable further opportunities to design novel all-oxide-on-semiconductor devices.
\end{abstract}
\maketitle

The magnetic-on-semiconductor heterostructures attract a lot of interest nowadays due to the vast opportunities they provide for designing novel functional spintronic devices for magnetic memory applications and bio-inspired computing \cite{prinz1990hybrid, ohno1999electrical, wolf2001spintronics,  yuasa2004giant,kent2015new,grollier2016spintronic,dieny2017perpendicular}. Placing a multiferroic layer with controllable magnetization/polarization in contact with a semiconductor adds the functionality of controlling optical, electronic and magnetic properties of the heterostructure by applied voltage \cite{scott2007data,gajek2007tunnel,ortega2015multifunctional,hu2017understanding}. One of the rare examples of material with spontaneous room-temperature magnetization and electric polarization is the metastable iron(III) oxide polymorph \eps{} \cite{gich2014multiferroic,ohkoshi2015nanometer,katayama2017chemical,xu2018origin}. Quite recently, the crystalline layers of \eps{} have been successfully synthesized on a number of oxide substrates \cite{gich2010epitaxial,gich2014multiferroic,thai2016stabilization,hamasaki2017crystal,corbellini2017epitaxially,viet2018specific} and GaN(0001) \cite{suturin2018tunable}. The structural and magnetic properties of the iron oxide films drastically depend on the composition of the neighboring buffer layer, the chosen substrate and the growth temperature. The feasibility to synthesize as much as four different iron oxide phases:  \eps{}, Fe$_3$O$_4$, $\alpha$-Fe$_2$O$_3$ and \gam{} on GaN(0001) by fine adjustment of growth parameters has been recently demonstrated \cite{suturin2018tunable}. It has been shown that stabilization of the \eps{} phase requires elevated growth temperature that leads to formation of a few nanometer-thick Ga-rich magnetically soft transition layer at the interface between the iron oxide film and the GaN substrate \cite{ukleev2018unveiling}. Later on, a very similar Ga/Fe substitution phenomena have been  observed in yttrium iron garnet (YIG) films grown at above 700$^\circ$C onto a gadolinium gallium garnet (GGG) \cite{suturin2018role}. Although $Pna2_1$ Ga-substituted epsilon-ferrite GaFeO$_3$ is isostructural to \eps{} \cite{abrahams1965crystal} and promotes further growth of the desired phase, its magnetic ordering temperature and coercivity field are somewhat lower than those of \eps{} \cite{katayama2017chemical}. This can potentially reduce the magnetoelectric and magnetooptical performance of the functional devices based on the \eps{}/ GaN heterostructures. 

In the present study, we have successfully introduced an epitaxial MgO buffer between the \eps{} and GaN layers to eliminate Ga migration into the iron oxide film. The resulting structural and magnetic properties of the fabricated heterostructure were probed by complementary x-ray diffraction (XRD), x-ray reflectometry (XRR), vibrating sample magnetometry (VSM), and polarized neutron reflectometry (PNR). An outcome of the epitaxial stabilization of \eps{} on the MgO buffer is a technological advantage that provides further opportunities to integrate the promising epsilon ferrite into epitaxial Fe \cite{goryunov1995magnetic,klaua2001growth,yuasa2004giant,raanaei2008structural,moubah2016discrete}, Fe$_3$O$_4$ \cite{anderson1997surface,margulies1997origin,gao1997growth,kim1997selective,voogt1999no}, $\alpha$-Fe$_2$O$_3$ \cite{gao1997growth,gao1997synthesis,kim1997selective} and \gam{} \cite{gao1997growth,voogt1999no} heterostructures and superlattices grown on MgO substrates.

\begin{figure}
\includegraphics[width=6.5cm]{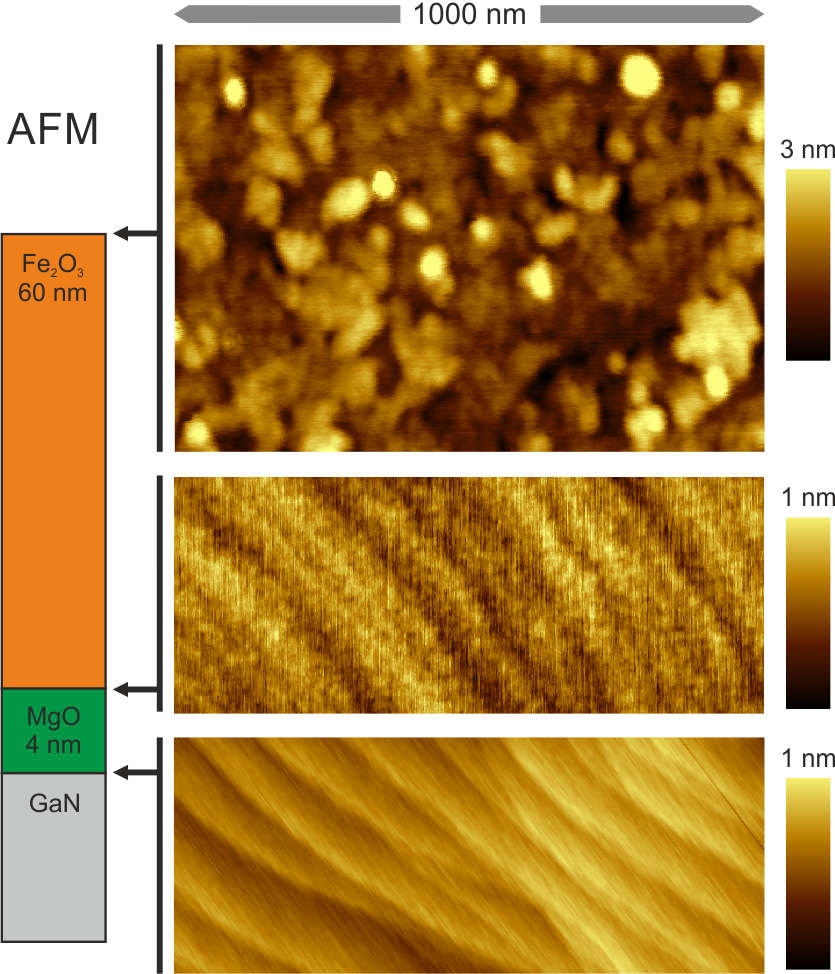}
\caption{(Color online) Atomic force microscopy images of the surface morphology at consecutive growth stages (from bottom to top): GaN, MgO/GaN and \eps{}/MgO/GaN.}
\label{afm}
\end{figure}

\begin{figure}
\includegraphics[width=6cm]{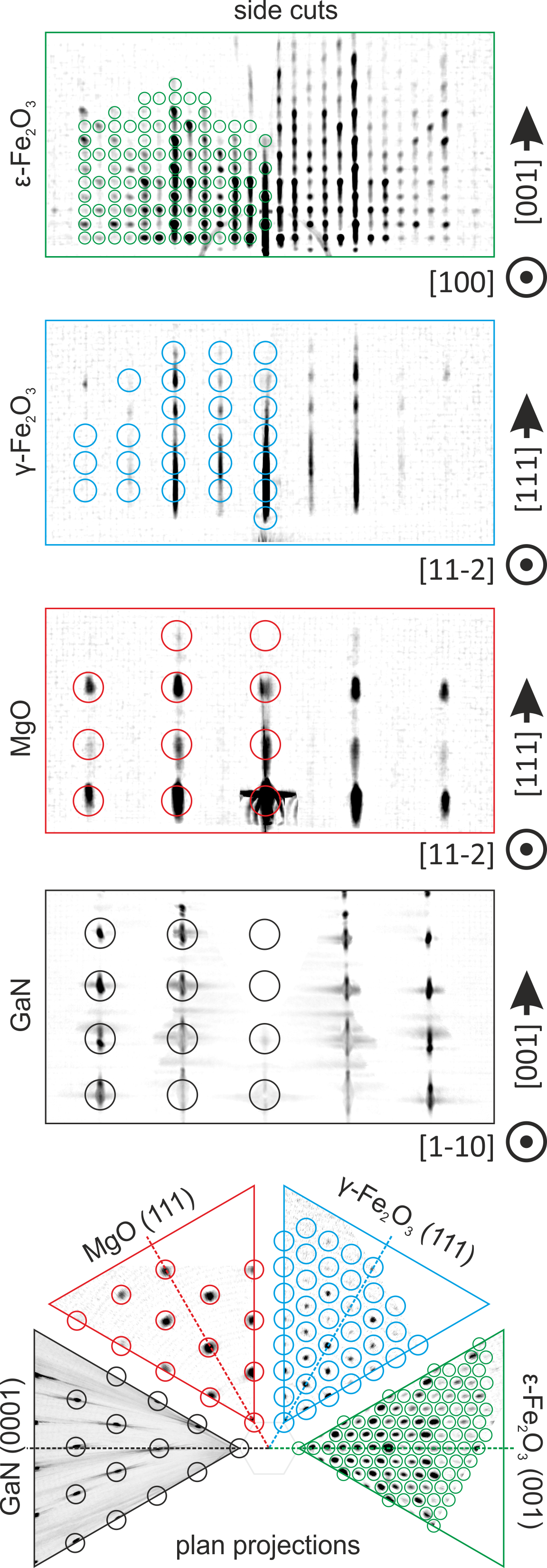}
\caption{(Color online) In-situ reflection high-energy electron diffraction maps obtained at consecutive growth stages: MgO/GaN, \gam{}/MgO/GaN and \eps{}/\gam{}/MgO/GaN. Shown in the same scale are the side cuts (top) and plan view projections (bottom) of the reciprocal space. The modeled reflection positions are shown with circles.}
\label{rheed}
\end{figure}

The substrates used in this work were commercial sapphire Al$_2$O$_3$ (0001) wafers with a 3\,$\mu$m-thick Ga terminated GaN (0001) layer grown on top by means of metalorganic vapour-phase epitaxy (MOVPE). The GaN surface showed a step-and-terrace surface morphology (Fig. \ref{afm}) as confirmed by atomic force microscopy (AFM). The oxide layers were grown by pulsed laser deposition (PLD) from MgO and Fe$_2$O$_3$ targets ablated using a KrF laser. The crystallinity and epitaxial relations of the grown layers were controlled by in-situ high energy electron diffraction (RHEED) reciprocal space 3D mapping. With this technique \cite{suturin2016} one obtains a 3D reciprocal space map from a sequence of conventional RHEED images taken during the azimuthal rotation of the sample. Thus obtained sequence of the closely spaced spherical cuts through the reciprocal space can be then compiled into a uniform 3D map and shown in the easy interpreted form of planar cuts and projections. The side cuts and plan views of the reciprocal space maps obtained at each growth stage are shown in the same scale in Fig. \ref{rheed}. The expected positions of the reciprocal lattice nodes are indicated with circles on the the left halves of the maps.

The 4\,nm thick MgO layer was deposited onto GaN in 0.02 mbar of oxygen at the substrate temperature of 800$^\circ$\,C. As confirmed by atomic force microscopy (Fig. \ref{afm}), the MgO coverage on GaN is smooth and sufficiently uniform to serve as a diffusion barrier. The epitaxial relations extracted from RHEED are as follows: GaN(0001) $||$ MgO(111); GaN[1-10] $||$ MgO$\pm$[11-2] (Fig. \ref{rheed}). The two possible MgO orientations arise due to the symmetry reduction occuring at the interface: from  GaN(0001) C$_6$ to MgO(111) C$_3$.  Reflections on the RHEED map of MgO are streaky corresponding to the semi-flat surface.

\begin{figure*}
\includegraphics[width=14cm]{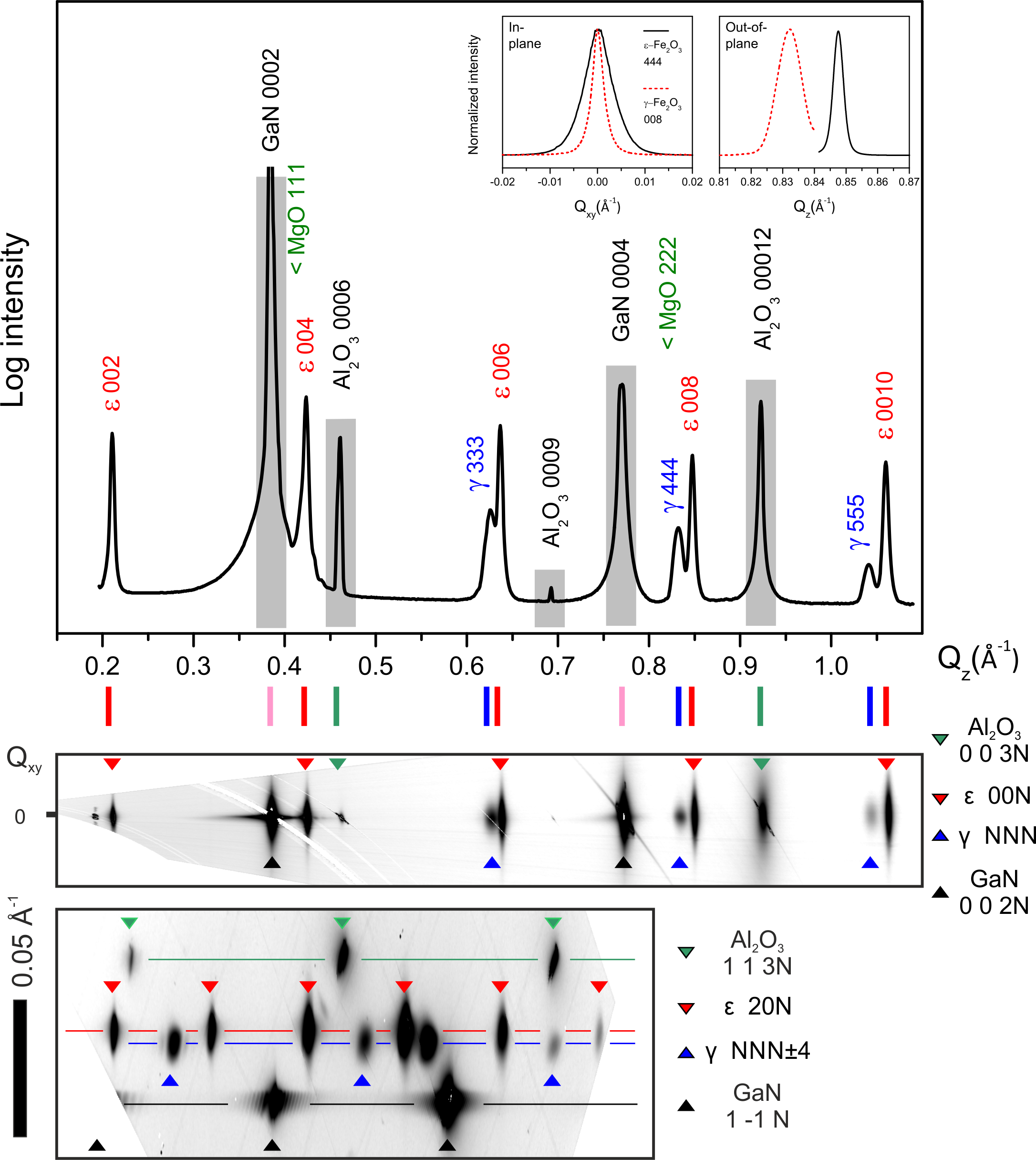}
\caption{(Color online) The XRD reciprocal space maps measured along the \eps{} 00N and 20N reflection chains in the \eps{}/\gam{}/MgO/GaN/Al$_2$O$_3$ sample. The specular intensity profile derived from the 00N map is shown on top. The insets show in-plane and out-of-plane widths of the \gam{} 444 and \eps{} 008 reflections. The reflections of each compound are labeled on the maps with triangles. }
\label{xrd}
\end{figure*}

A 60\,nm thick iron oxide layer was grown onto the surface of MgO(111) in 0.2 mbar of oxygen at the substrate temperature of 800$^\circ$\,C following the approach described in our previous report \cite{suturin2018tunable}. It was discovered that unlike when grown directly on GaN, the iron oxide layer on MgO nucleates in gamma rather than in epsilon phase. Upon deposition of 3-5\,nm of iron oxide, the RHEED reciprocal space maps start showing a distinct $2\times2$ pattern of streaks characteristic for the spinel \gam{} lattice (Fig. \ref{rheed}) oriented with the [111] axis perpendicular to the surface and the [11-2] axis parallel to MgO [11-2] and GaN[1-10]. The diffraction map remains streaky corresponding to the still flat surface.

The preference of the \gam{} over \eps{} is naturally related to the cubic symmetry of both lattices. The phase choice mechanisms for the Fe$_2$O$_3$ / MgO(111) system might be similar to those of the Fe$_2$O$_3$ / MgO (001) system where \gam{} is known to be the dominant phase \cite{gao1997growth,huang2013epitaxial,sun2014effect}. It is noteworthy that a thin $\gamma$-like transition layer was also observed during the nucleation of $\alpha$- and \eps{} directly on GaN \cite{suturin2018tunable}. Though the diffraction patterns of that layer bore resemblance to FeO, the spacing between the adjacent (111) layers of oxygen was very similar to \gam{}. 

When the total thickness of the iron oxide reaches about 10\,nm, the $2\times2$ streak pattern gets gradually replaced by the $6\times1$ streak pattern which is an unmistakable fingerprint of the \eps{} phase. This pattern persists until the growth is stopped at 60\,nm of the iron oxide total thickness (Fig. \ref{rheed}). The pattern is dotty rather than streaky in agreement with the few nm surface roughness measured by AFM (Fig. \ref{afm}). The \eps{} lattice is oriented with the polar [001] axis perpendicular to the surface and the easy magnetization [100] axis parallel to the one of the three equivalent GaN [1-10] in-plane directions resulting in three crystallographic domains at $120 ^\circ$ to each other. It is essential that the growth temperature at this stage is no less than 800$^\circ$C otherwise nucleation of \eps{} phase does not occur.

To accurately study the crystal structure of the film volume we have applied X-ray diffraction in addition to the surface sensitive RHEED. The XRD measurements were carried out at the BL-3A beamline, KEK Photon Factory (Tsukuba, Japan). The 3D reciprocal space maps were compiled from a series of diffraction patterns taken with a Pilatus 100K two-dimensional detector during a multi-angle rotation performed on a standard 4-circle Euler diffractometer. The linear and planar cuts through the 3D maps obtained across the reciprocal space specular region are shown in Fig. \ref{xrd}. The series of \eps{} 002$\cdot$N and \gam{} 111$\cdot$N reflections are easily identifiable in addition to the reflections of the underlying Al$_2$O$_3$ and GaN. We do not observe distinctly the reflections of MgO as they considerably overlap with those of \gam{}. Moreover the MgO layer is 15 times thinner than Fe$_2$O$_3$ and has about 1.5 times lower scattering length density for x-rays. 

The derived out-of-plane lattice constant of epsilon ferrite $c=9.43$\,{\AA} is in agreement with our earlier studies of \eps{} / GaN \cite{suturin2018tunable}. The (111) interplane distance in \gam{} is in agreement with the bulk lattice constant of \gam{} $a=8.33$\,\AA. The in-plane lattice arrangement becomes clear from the analysis of the reciprocal space region containing the off-specular \eps{} 20N reflections. The \eps{} lattice shows a 1\% in-plane expansion towards $a=5.14$\,\AA\, and $b=8.86$\,\AA. The \gam{} lattice shows a 1.5\,\% lattice expansion towards the equivalent cubic lattice constant of $a=8.47$\,\AA. The in-plane expansion is not surprising taking into account the fact that the in-plane periodicity in GaN is about 8.5\%  larger than in Fe$_2$O$_3$ \cite{suturin2018tunable}. 
The observed in-plane and out-of-plane reflection widths may be used to judge on the strain relaxation and minimal crystallographic domain size in the grown films. The strain relaxation if present would involve a distribution of lattice parameters in the system and would cause reflection broadening that is proportional to the magnitude of the wave vector $Q_z$. Even if such a broadening is present in our system, it is below the experimental resolution as all the observed reflections are of the same shape and width. Such effect can be attributed to the finite size of the coherent crystallographic domains within the crystal lattice and is typical for the nanostructured samples. Measuring the in-plane and out-of-plane reflection widths (see the insets in Fig. \ref{xrd}) one can conclude that the minimal coherent domains of \eps{} are shaped as (width$\times$height) 14\,nm $\times$ 35\,nm columns (in agreement with Ref. \cite{ukleev2018unveiling}) while those of \gam{} look like 33\,nm\,$\times$\,10\,nm disks. The reduced coherent thickness of \eps{} film suggests that a transition layer with a mixed lattice structure exist at the \gam{}/\eps{} interface. The lateral coherence between the the adjacent nucleation sites is substantially reduced because the surface cell of the iron oxides is larger than that of MgO. Compared to \gam{} the coherent domain of \eps{} is smaller because of the larger surface cell and the lack of the C$_3$ symmetry. Thus the antiphase boundaries are formed more frequently in \eps{}.

\begin{figure}
\includegraphics[width=8.5cm]{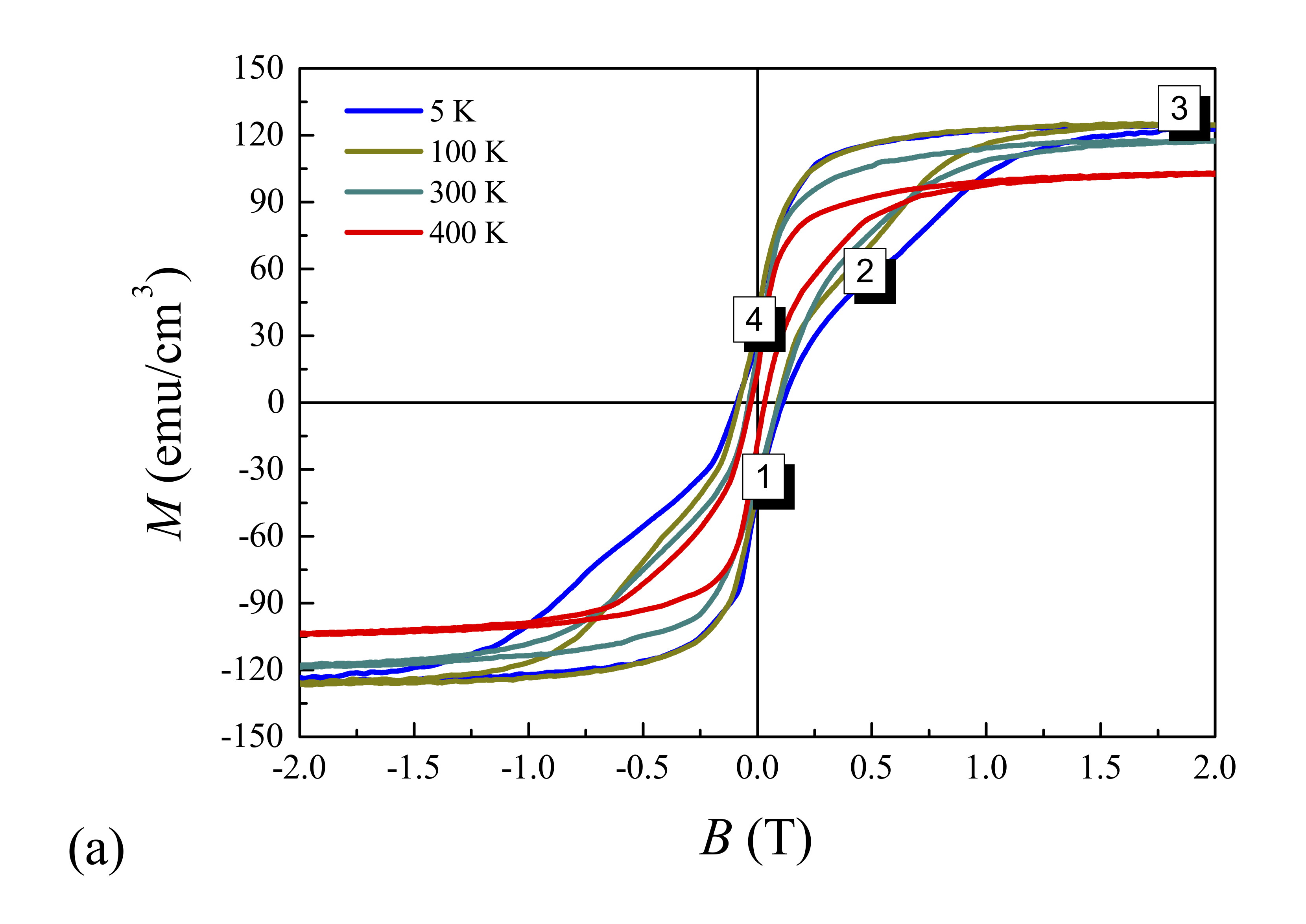}
\includegraphics[width=8.5cm]{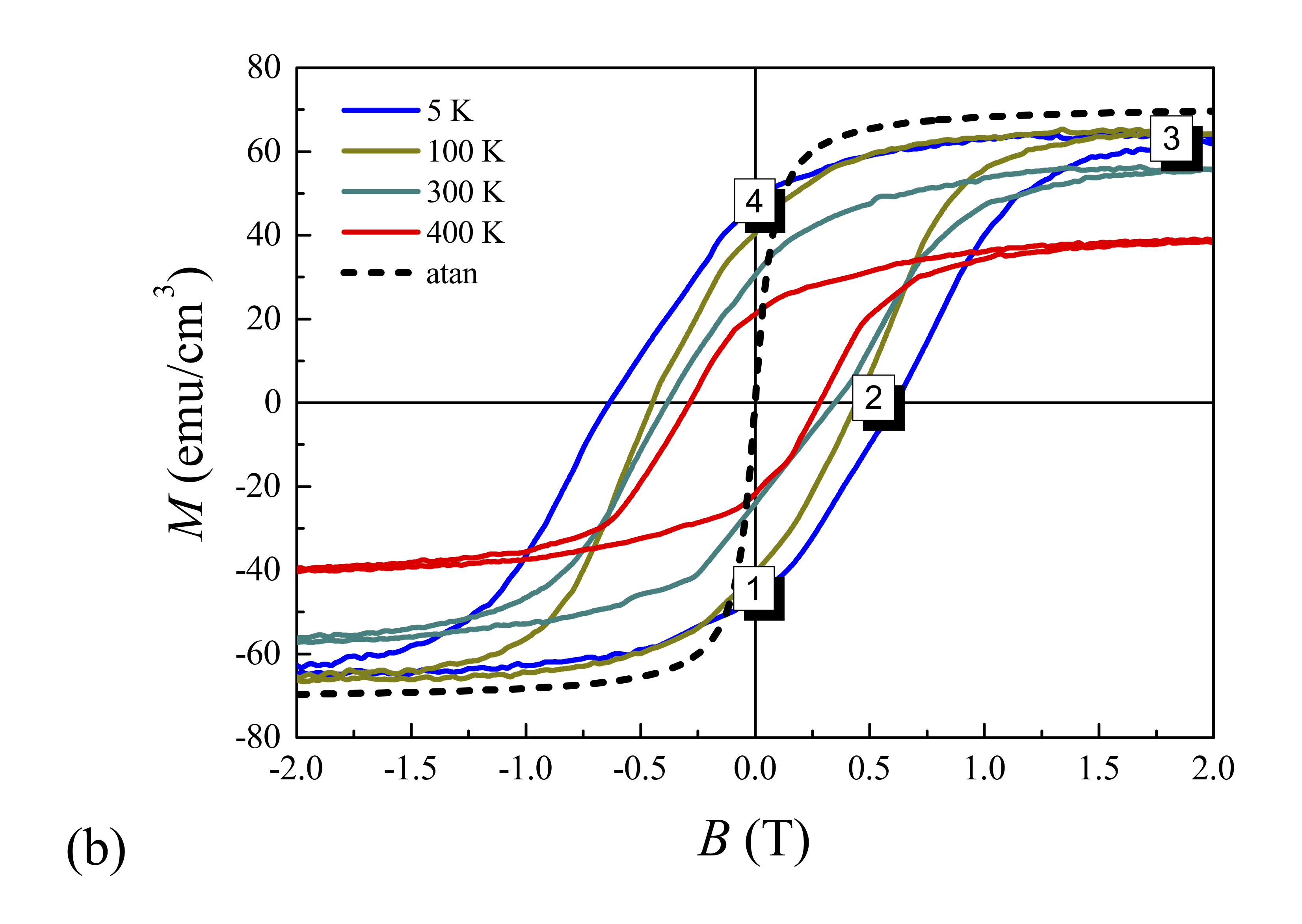}
\caption{(Color online) In-plane hysteresis $M(B)$ curves of 70\,nm-thick \eps{}/MgO film measured at 5-400\,K. Shown are curves (a) as measured and (b) decomposed to the hard and soft components. To express the magnetization in emu/cm$^3$ the curves in (a) are normalized to the expected film thickness of 70\,nm. The hard and soft component curves in (b) are normalized to the thicknesses of 60\,nm corresponding to the thickness of \eps{} layer and 70\,nm corresponding to the total thickness of the sample.}
\label{squid}
\end{figure}

The magnetometry measurements were carried out using a Quantum Design PPMS vibrating-sample magnetometer (VSM). The magnetic field was applied in the sample plane along the [100] easy magnetization axis of one of the three \eps{} domains. Fig. \ref{squid} shows the hysteresis loops measured in the temperature range of 5-400\,K and corrected for the linear diamagnetic contribution of the substrate. The observed values of saturation magnetization were about 130\,emu/cm$^3$ at $T\,=\,5$\,K and 100\,emu/cm$^3$ at $T\,=\,400$\,K which is consistent with what was reported for \eps{} nanoparticles \cite{jin2004giant} and \eps{} thin film grown on SrTiO$_3$ (STO) \cite{gich2014multiferroic}, YSZ \cite{corbellini2017epitaxially,knivzek2018spin} and GaN\cite{ukleev2018unveiling}, and predicted from ab-initio calculations \cite{xu2018origin}. 

\begin{figure*}
\includegraphics[width=16.5cm]{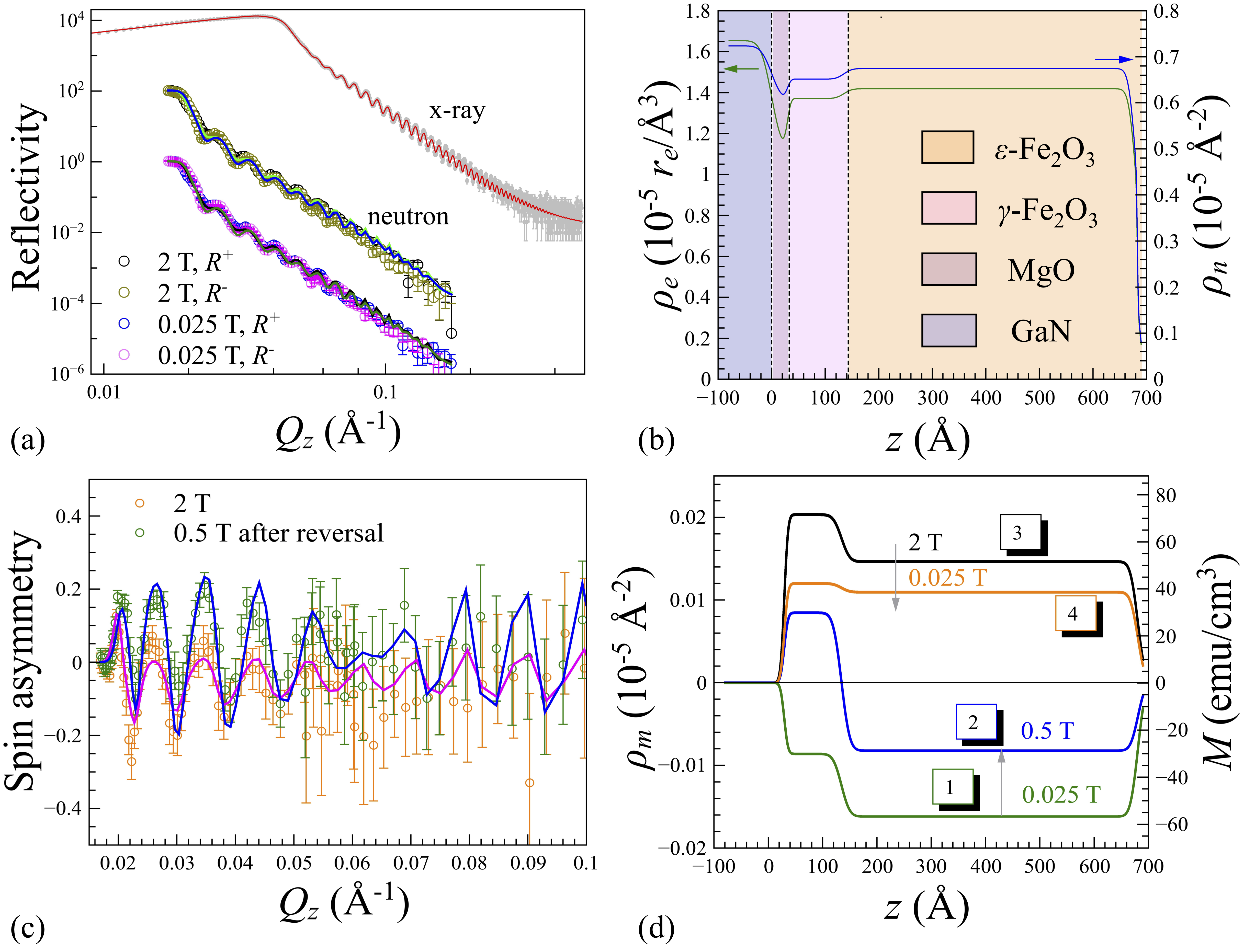}
\caption{(Color online) (a) Measured (symbols) and fitted (solid lines) x-ray and neutron reflectivity curves as a function of momentum transfer ($Q_z$) on a logarithmic scale. The curves are shifted along vertical axis for clarity. (b) X-ray scattering length density (SLD) $\rho_e$ (green line), and neutron nuclear SLD $\rho_n$ (red line) of \eps{}/MgO/GaN film as a function of the distance from the GaN layer surface ($z$) obtained from the fitting routine. X-ray SLD $\rho_e$ is given in the units of the classical electron radius $r_e=2.81794...\times 10^{-15}$\,m. (c) PNR spin-asymmetry ratio ($R^+ - R^-)/(R^+ + R^-)$ at applied magnetic field $B=2$\,T and $B=0.5$\,T after magnetization reversal obtained from experimental data (symbols) and fitted models (solid curves). (d) Neutron magnetic SLD $\rho_m$ profile at $B=2$\,T, $B=0.025$\,T before and at $B=0.025$\,T, $B=0.5$\,T after magnetization reversal, corresponding to the characteristic points ($1-4$) of the $M(B)$ loop shown in Fig.\ref{squid}.}
\label{xrrpnr}
\end{figure*}

The wasp-waist magnetization loops shown in Fig. \ref{squid}a are typical for \eps{} films and nanoparticles and can be qualitatively decomposed to hard and soft component loops (Fig. \ref{squid}b) by subtracting $2M_{soft}/\pi \cdot$ arctan($B$/$B_{soft}$) function with temperature-independent $M_{soft}\,=\,71$ emu/cm$^3$ and $B_{soft}$\,=\,62\,mT. These parameters were unambiguously derived from manual optimization aimed at making the remaining hard component smooth and monotonous in the vicinity of zero magnetic field.

The value of $M_{soft}\,=\,71$ emu/cm$^3$ observed for the soft magnetic component is in general agreement with the presence of \gam{} sublayer buried below the main layer of \eps{} as observed by XRD, RHEED and PNR. The magnetization plotted in Fig. \ref{squid}b is normalized to the total film thickness of 70 nm. Taking into account the reported values of $M_s$=300-400 emu/cm$^3$ for \gam{} / MgO, the soft loop can be attributed to a layer of \gam{} having thickness of 12-14 nm. This is comparable though slightly higher than the thickness estimated from RHEED and PNR (see the details below).

The hard component hysteresis loops show a large saturation field of 1.2-1.8\,T characteristic of \eps{}. The coercive field gradually increases as the sample is cooled down - from 0.27\,T at 400\,K to 0.66\,T at 5\,K. The loop shape is typical for the system with three uniaxial domains at 120 deg to each other. At saturation the magnetization is collinear to the field in all three domains $M_s^{sum}=3\cdot M_s$. From saturation to zero field the magnetization gradually decreases to $2/3\cdot M_s^{sum}$ as the the magnetization in the two non collinear domains returns to the equilibrium state at 120\,deg to the field. From this state the magnetization reversal is gradually completed towards the negative saturation. Notably, the magnetic phase transition to an incommensurate state that is often observed in \eps{} nanoparticles, as dramatic shrinkage of the loop at $T\approx100-150$\,K \cite{gich2006high,tseng2009nonzero,garcia2017unveiling,ohkoshi2017large}, has not been observed in \eps{} films - neither on GaN nor on the other substrates. The absence of a sharp phase transition in films can be caused by the variation of the magnetic properties across the film depth. Thus, a temperature-dependent investigation of the depth resolved magnetic structure of \eps{} films by neutron or resonant x-ray diffraction is highly desired to address this issue. 

The XRR measurement was performed on the Panalytical X'Pert PRO x-ray diffractometer at room temperature using Cu $K_\alpha$ (1.5406\,\AA) radiation to determine the electron scattering length density (SLD) profile $\rho_e$ of the film as a function of distance from the GaN surface $z$. The specular reflectance was measured in the range of incident angles between 0.5 to 3.5\,degrees covering the $Q_z$ range from 0.075 to 0.5\,\AA$^{-1}$.

The neutron reflectometry experiments were performed at the D17 setup \cite{saerbeck2018recent,illdata2018} (ILL, Grenoble, France) in polarized time-of-flight mode. Sample temperature and magnetic field were controlled by an Oxford Instruments 7\,T vertical field cryomagnet equipped with single-crystalline sapphire windows. Neutrons with wavelengths of $4-16$\,\AA~were used to ensure the constant polarization of $P_0\,>\,99\%$. Three different incident angles (0.8, 1.5 and 3.7\,degrees) were chosen to access the $Q_z$ range from 0.017 to 0.17\,\AA$^{-1}$. Intensity of the reflected beam was collected by two-dimensional $^3$He position-sensitive detector. The data was integrated using a method taking into account the sample curvature or beam divergence \cite{saerbeck2018recent,cubitt2015improved}. Non-spin-flip reflectivities $R^+$ and $R^-$, where +(-) denotes the incident neutron spin alignment parallel (antiparallel) to the direction of applied magnetic field, were acquired without polarization analysis. The detailed description of the reflectometry techniques can be found elsewhere \cite{ankner1999polarized,zabel2008polarized}.

Figure \ref{xrrpnr}a shows x-ray reflectivity (room temperature) and neutron reflectivity ($T=5$\,K) curves plotted as a function of momentum transfer $Q_z$. The neutron reflectivity curves were measured at the characteristic characteristic points of the $M(B)$ loop marked as ($1-4$) in Fig.\ref{squid}. The PNR curves shown in Fig. \ref{xrrpnr}a are measured in applied magnetic fields of $B=0.025$\,T (state 1 in remanence) and $B=2$\,T (state 3 in saturation). The XRR and PNR curves were simultaneously fitted using GenX software \cite{bjorck2007genx}. The simplest model, for which the fitting routine converges, corresponds to a stack consisting of the GaN substrate, the MgO buffer, the transition iron oxide layer with an unspecified density and the main \eps{} layer. The depth-profiles of the x-ray ($\rho_e$) and nuclear neutron ($\rho_n$) scattering length densities (SLDs) extracted from the refined model are shown in Fig. \ref{xrrpnr}b. The profiles reflect the chemical composition and density of the layers as well as the structural roughness of the interfaces. The root mean square (RMS) roughness of all the interfaces is below 15\,\AA. Notably, we observe the transition layer at the iron oxide/MgO interface with thickness of $105\pm10$\,\AA~and reduced x-ray and neutron nuclear SLDs compared to the main \eps{} volume of the film. This looks natural as \gam{} having the same chemical formula as  \eps{} is by 3.4\,\% less dense due to the presence of iron vacancies in the inverted spinel structure. The comparably low SLD of the MgO layer gives a few nm wide reduction of $\rho_e$ and $\rho_n$ located on the SLD profile at $z=0$.

The magnetization profile of the heterostructure is encoded in the dependence of the spin-asymmetry ratio ($R^+$-$R^-$)/($R^+$+$R^-$) on $Q_z$. Fitting it against the model gives the depth profile of the magnetic contribution to the neutron SLD $\rho_m$ (\AA$^{-2}$) which can be converted to magnetization $M$ (emu/cm$^3$) using the following formula: $M=\,3505\cdot 10^5 \cdot \rho_m$ \cite{zhu2005modern}. The measured and fitted spin-asymmetry ratios are shown in Fig. \ref{xrrpnr}c for the two magnetic states 2 and 3 on the lower branch of the hysteresis loop (see Fig. \ref{squid}): with partially switched magnetization ($B=+0.5$\,T) and in full saturation ($B=+2$\,T). The fitted model suggests that the iron oxide film is divided into two magnetically different sub-systems: the main \eps{} layer with a saturation magnetization of $M_{s1} \approx 56$\,emu/cm$^3$ and an interfacial layer with $M_{s2}\approx70$\,emu/cm$^3$ (Fig. \ref{xrrpnr}d). Using the PNR data obtained at 5\,K we are able to track the magnetization behavior of individual sublayers as the system is magnetized from the negative remanence (state 1) to full saturation (state 3) and back to the positive remanence (state 4). As shown in (Fig. \ref{xrrpnr}d) the magnetization of the softer interface layer is switched between $B=0.025$\,T (state 1) and $B=0.5$\,T (state 2) and reaches saturation of 70\,emu/cm$^3$ at $B=2$\,T. The magnetization of the much harder \eps{} layer switches somewhere between $B=0.5$\,T (state 2) and $B=2$\,T (state 3). As the magnetically hard component of the hysteresis loop is not completely closed in the maximum applied positive of 2\,T (Fig. \ref{squid}b), the PNR curves measured at $B=2$\,T (state 3) and $B=0.025$\,T (state 4) belong to the minor branch of the hysteresis. Magnetization of 56\,emu/cm$^3$ is found at $B=2$\,T, which is slightly smaller that the saturation moment. Going back to positive remanence of the minor loop (state 4), the magnetization of both interface and bulk layers start slowly decreasing (faster for the interface layer).

Sequential switching of interface $\gamma$- and main $\varepsilon$- layers in principle reflects a step-like shape of the hysteresis loops observed by VSM magnetometry (Fig. \ref{squid}). It must be noted that the maximum magnetization for \eps{} layer derived from PNR is about twice lower than the highest reported values for \eps{} but in good agreement with the maximum magnetization observed in the decomposed VSM loop shown in Fig. \ref{squid}b. The maximum magnetization of the \gam{} layer derived from PNR is about 5 times lower than the expected 300-400\,emu/cm$^3$ reported for \gam{}/MgO layers \cite{gao1997growth,huang2013epitaxial,sun2014effect}, and cannot completely explain the soft-magnetic component observed by VSM. Magnetic degradation of the transition \gam{} layer can be possibly explained by the size effect \cite{orna2010origin}, epitaxial strain \cite{yang2010strain,bertinshaw2014element,gibert2015interfacial} or large number of the antiphase boundaries \cite{rigato2007strain,ramos2009artificial} between the nano-columns in the plane of the layer and at the interface with main \eps{} film.

The much higher magnetization of the soft magnetic component observed in VSM suggests that another soft magnetic phase is likely present in the sample that cannot be distinguished in the PNR experiment. Similar effect was also observed in \eps{} grown directly on GaN \cite{ukleev2018unveiling}. The most plausible candidates are homogeneously distributed minor fractions of polycrystalline \gam{} and Fe$_3$O$_4$  \cite{jin2005formation,lopez2016growth,corbellini2017effect} not pronounced in XRD data. Again, one must also take into account the columnar structure of the \eps{} films containing considerable concentration of the antiphase boundaries. As was pointed out in Ref. \cite{sofin2011anomalous} the antiphase boundaries in iron oxides may account for the soft magnetic behavior. The magnetic moments located in minor phase fractions of small volume, or at the antiphase boundaries in the sample plane that cannot be resolved with PNR, which is a laterally averaging technique, because the disordered moments at boundaries and minor phase fractions are highly diluted, but integrated into the magnetization measured by VSM. We suggest that the deposition of small ($\mu$m-scale) iron particulates ejected from the PLD target is the most plausible scenario, that have been also observed for other PLD films \cite{haindl2016situ,zhai2018weak,grant2018particulate}.

In conclusion, we have demonstrated the possibility to epitaxially grow single crystal \eps{} thin film on MgO(111) surface by pulsed laser deposition. In contrast to the previously investigated non-buffered \eps{}/GaN(0001) system, where the interfacial GaFeO$_3$ magnetically degraded layer was reported to form due to Ga diffusion\cite{ukleev2018unveiling} from GaN, the \eps{} / MgO / GaN system has advantage of exploiting the diffusion blocking MgO barrier. Though formation of the orthorhombic GaFeO$_3$ was supposed earlier to be a potential trigger of the nucleation of the isostructural \eps{}, the present work demonstrates that the growth of single crystalline uniform films of epsilon ferrite by pulsed laser deposition is possible even without the aid of Ga. Still the aid of Ga seems important as on GaN the \eps{} layer could be nucleated with a transition layer of few angstrom thickness while on MgO the growth of \eps{} film is preceded by nucleation of a 10\,nm thick layer of another iron oxide phase. A complimentary combination of electron and x-ray diffraction, x-ray reflectometry and polarized neutron reflectometry techniques allowed unambiguous identification of this phase as $P4_132$ ($P4_332$) cubic \gam{}. This phase is known to show magnetoelectric functionality \cite{cheng2016enhancements} and spin Seebeck effect \cite{jimenez2017spin} and can enable further opportunities to design the novel all-oxide heterostructure magnetoelectric and spin caloritronic devices.

We are grateful to Institut Laue-Langevin for provided neutron and x-ray reflectometry beamtime (proposal No.: 5-54-244 \cite{illdata2018}). Synchrotron x-ray diffraction experiment was performed at KEK Photon Factory as a part of the proposal No. 2018G688. We thank Dr. Tian Shang for the assistance with magnetization measurements. The part of the study related to PNR and XRR was partially supported by SNF Sinergia CRSII5-171003 NanoSkyrmionics. The part of the study related to growth technology and diffraction studies was supported by Russian Foundation for Basic Research grant \textnumero 18-02-00789. The open access fee was covered by FILL2030, a European Union project within the European   Commission's Horizon 2020 Research and Innovation programme under grant agreement No. 731096.

\end{document}